\documentclass{article}
\usepackage{mathptmx}       
\usepackage{amsmath, amsfonts, amssymb} 
\usepackage{PRIMEarxiv}

\usepackage[utf8]{inputenc} 
\usepackage[T1]{fontenc}    
\usepackage{hyperref}       
\usepackage{url}            
\usepackage{booktabs}       
\usepackage{amsfonts}       
\usepackage{nicefrac}       
\usepackage{microtype}      
\usepackage{lipsum}         
\usepackage{fancyhdr}       
\usepackage{graphicx}       
\graphicspath{{media/}}     

\usepackage{cases}
\usepackage{epsfig}         

\usepackage{fullpage, lineno, setspace, nicefrac, subcaption, caption}
\usepackage{mathrsfs}
\usepackage{tikz}
\usetikzlibrary{automata,arrows,positioning,calc}
\usepackage[mathscr]{eucal}
\usepackage{xcolor}
\usepackage{color}
\DeclareGraphicsExtensions{.png,.jpeg,.jpg,.pdf}
\graphicspath{{./Figures/}}

\newtheorem{theorem}{Theorem}[section]

\definecolor{darkgreen}{rgb}{0,0.5,0}

\newcommand{\set}[1]{\left\{ #1 \right\}}
\newcommand{\Def}{\overset{\textbf{def}}{=}}
\newcommand{\RR}{\mathbb{R}}

\newcommand{\EE}{\mathbb{E}}

\newcommand{\rhomax}{\rho_{\text{max}}}
\newcommand{\fpartial}[2]{\frac{\partial #1}{\partial #2}}

\newcommand{\totvar}[2]{\underset{#2}{\mathrm{TV}}\left(#1 \right)}

\newcommand{\BV}{\mathbf{BV}}

\DeclareMathOperator{\loc}{loc}

\newcommand{\nnorm}[1]{{\left\vert\kern-0.25ex\left\vert\kern-0.25ex\left\vert #1 
    \right\vert\kern-0.25ex\right\vert\kern-0.25ex\right\vert}}

\title{\LARGE 
Reinforcement learning-based adaptive speed controllers in mixed autonomy condition 
}

\author{H. Wang\textsuperscript{1}, H. Nick Zinat Matin\textsuperscript{1}, M. L. {Delle Monache}\textsuperscript{1}
\thanks{$^{1}$H. Wang, H. Nick Zinat Matin, and M. L. Delle Monache are with the Department of Civil and Environmental Engineering,         University of California, Berkeley, USA
        {\tt\small hanw@berkeley.edu, h-matin@berkeley.edu, mldellemonache@berkeley.edu }}%
}

\begin{document}

\maketitle

\begin{abstract}
The integration of Automated Vehicles (AVs) into traffic flow holds the potential to significantly improve traffic congestion by enabling AVs to function as actuators within the flow. This paper introduces an adaptive speed controller tailored for scenarios of mixed autonomy, where AVs interact with human-driven vehicles. We model the traffic dynamics using a system of strongly coupled Partial and Ordinary Differential Equations (PDE-ODE), with the PDE capturing the general flow of human-driven traffic and the ODE characterizing the trajectory of the AVs. A speed policy for AVs is derived using a Reinforcement Learning (RL) algorithm structured within an Actor-Critic (AC) framework. This algorithm interacts with the PDE-ODE model to optimize the AV control policy. Numerical simulations are presented to demonstrate the controller's impact on traffic patterns, showing the potential of AVs to improve traffic flow and reduce congestion.
\end{abstract}

\section{INTRODUCTION}




The emergence of technological advances in transportation systems holds the promise of improving traffic congestion and reducing fuel/energy consumption and consequently reducing pollution. In recent years, research has focused on the use of Autonomous Vehicles (AVs) in traffic flow as controllers.

Some of the works in the control framework include \cite{talebpour2016influence,
piacentini2018traffic}, in the machine learning framework ~\cite{Wu2018} and real world experiments~\cite{stern2018dissipation}. 
In Liard et al. \cite{liard2021pde} the AVs' adaptive speed is cast as the solution to an optimal control problem with the objective of minimizing fuel consumption. They show the existence of a solution to the optimal control problem.
All these works show that Lagrangian control is a viable alternative to the classical approach to traffic control involving Variable Speed Limit (VSL) ~\cite{ Hegyi2005,  delle2017traffic}, without requiring dedicated infrastructure. 

More recently, Reinforcement Learning (RL) has emerged as a powerful tool for optimizing decision-making processes in complex, dynamic environments. In the context of traffic control, RL algorithms have been applied to a variety of problems, ranging from signal control to route planning. Few works have focused on the control of AVs through RL. Krauss et al. \cite{krauss1998microscopic} were among the first to explore the use of RL for controlling AVs, where they applied a microscopic traffic flow model to train vehicles to optimize their speed profiles. More recently, Shalev-Shwartz et al. \cite{shalev2016safe} proposed a safe, multi-agent RL framework for autonomous driving, which not only learns efficient driving policies but also ensures safety constraints are met. Wu et al. \cite{wu2017flow} introduced  a computational framework for deep RL and control experiments for traffic optimization, which includes AVs as part of the traffic flow. Their work demonstrates the potential of RL to reduce congestion and improve traffic efficiency by controlling a small fraction of the vehicles on the road.
\begin{figure}
    \centering
    \includegraphics[width=0.77\linewidth]{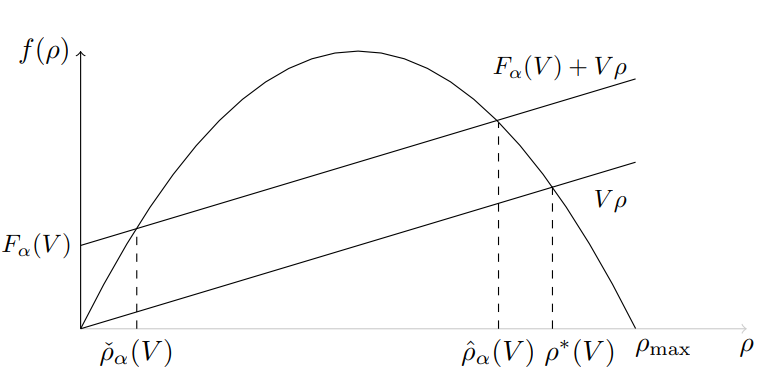}
    \caption{Illustration of the FD and the locations of $\check \rho(V)$ and $\hat \rho(V)$ for each $V \in [0, V_{\max}]$. The solutions above the line $F_\alpha(V) + V \rho$ do not satisfy flux constraint \eqref{E:flux_constraint}.}
    \label{fig:FD}
\end{figure}

\textbf{Contribution.} In this paper, we focus on employing an RL method to control AVs immersed in bulk human traffic to regulate congestion. The goal is to use autonomous vehicles as actuators in a PDE-ODE model framework where the bulk flow of the traffic is captured by PDE and an AV trajectory is represented 
 by an ODE. In this paper, in particular, we show that an improvement in traffic metrics, such as minimum flow rate and average velocity of the traffic, can be achieved by designing an RL algorithm such that the corresponding reward function is updated in interaction with the PDE-ODE model. \\
The paper organization is as follows: Section \ref{sec:math_model} defines the problem and demonstrates how to find solutions to the PDE-ODE problem. In addition, in this section, we will discuss some of the analytical results from the control theory which will be used later to show the validity of our proposed method.  
Section \ref{sec:control_design} details the proposed controller and the connection with PDE-ODE will be sketched. Then, numerical results will be elaborated in Section \ref{S:numerical}. We finalize the paper by discussing the results and possible extensions in Section \ref{sec:conclusion}. 

\section{Mathematical Model} \label{sec:math_model}



In this paper, we consider a PDE-ODE model of the form
\begin{subnumcases}{}\label{eq:model}
    \rho_t + \fpartial{}{x}[f(\rho)]= 0   \label{E:conservation}\\
   \rho(0, x) = \rho_\circ(x) \label{E:initial_conserv}\\
    f(\rho(t, y_i(t)) - \dot y_i(t) \rho(t, y_i(t)) \le F_\alpha(\dot y_i(t))  \label{E:flux_constraint}\\
    \dot y_i(t) = \min \set{V_i(t), v(\rho(t, y_i(t)+))}  \label{E:AV_traj}\\
    y_i(0) = y_{i,\circ}.  \label{E:initial}
\end{subnumcases} 
where, $f(\rho) = \rho v(\rho)$ is the flux function, $v(\rho) = V_{\max} \left(1 - \frac{\rho}{\rho_{\max}} \right)$ is the average traffic velocity, $V_{\max}$ the maximum velocity, $\rho$ is the average traffic density, $\rho_{\max}$ is the maximum density, and $t \mapsto V_i(t)$ denotes the maximum desired speed of the $i$-th AV. Equations \eqref{E:conservation} and \eqref{E:initial_conserv} describe the evolution in time of the traffic density (also known as the LWR model, \cite{lighthill1955kinematic, richards1956shock}) and Equations \eqref{E:AV_traj} and \eqref{E:initial} describe the AV dynamics. Here, $\rho(t, y(t)+) \Def \lim_{x \searrow y(t)} \rho(t, x)$ states that the velocity is only influenced by the downstream density. Finally, the inequality \eqref{E:flux_constraint} captures the decrease in the flux in the Lagrangian coordinates by a factor $\alpha \in (0, 1)$ as a result of slow-moving AV; see \cite{delle2014scalar, liard2021entropic, matin2023existence} for more details. We define,
\begin{equation}
F_\alpha(\dot y_i) \Def \max_{\rho \in [0, \rho_{\max}]} (f(\rho) - \dot y_i \rho).\end{equation}
The generalization of the Cauchy problem \eqref{E:conservation}-\eqref{E:initial} to the case of variable speed limit and space dependent flux has been studied in \cite{matin2023existence} for the existence of weak solution. 
In this paper for simplicity, we consider a single AV and $V_i = V$. It should be noted the generalization of the results to several AVs is possible, see for example \cite{garavello2020multiscale, goatin2023interacting}. 
For each fixed $V \in [0, V_{\max}]$ we define $\check \rho(V)$ and $\hat \rho(V)$ as the intersection points of $F_\alpha(V) + \rho V$ with the fundamental diagram, such that $\check \rho (V) < \hat \rho(V)$; see Figure \ref{fig:FD}. 

\begin{theorem}\label{T:existence}
    Let $\rho_\circ \in \BV(\RR;[0 , \rho_{\max}])$, where $\BV$ represents the space of bounded variations. Then, the Cauchy problem \eqref{E:conservation}-\eqref{E:initial} has a weak entropy solution 
    $(\rho, y) \in C(\RR_+; L^1 \cap \BV(\RR; [0, \rhomax])) \times W^{1,1}_{\loc}(\RR_+ ; \RR)$ in Kruzkov sense \cite{kruvzkov1970first}.
\end{theorem}
The main idea of the proof is similar to the classical proof through the wavefront tracking scheme and by defining approximate solution $(\rho^{(n)}, y_n)$ and then letting $n \to \infty$. However, in addition to such approximation, in this theorem the function $V$ will also be approximated by a piecewise constant function. The proof of this theorem as well as the weak entropy definition for this problem can be found in \cite{garavello2020multiscale} for the single AV and in \cite{liard2020optimal, liard2021pde} for multiple AVs. 
\section{Controller Design}\label{sec:control_design}
\begin{figure}
    \centering
    \includegraphics[width=0.9\linewidth]{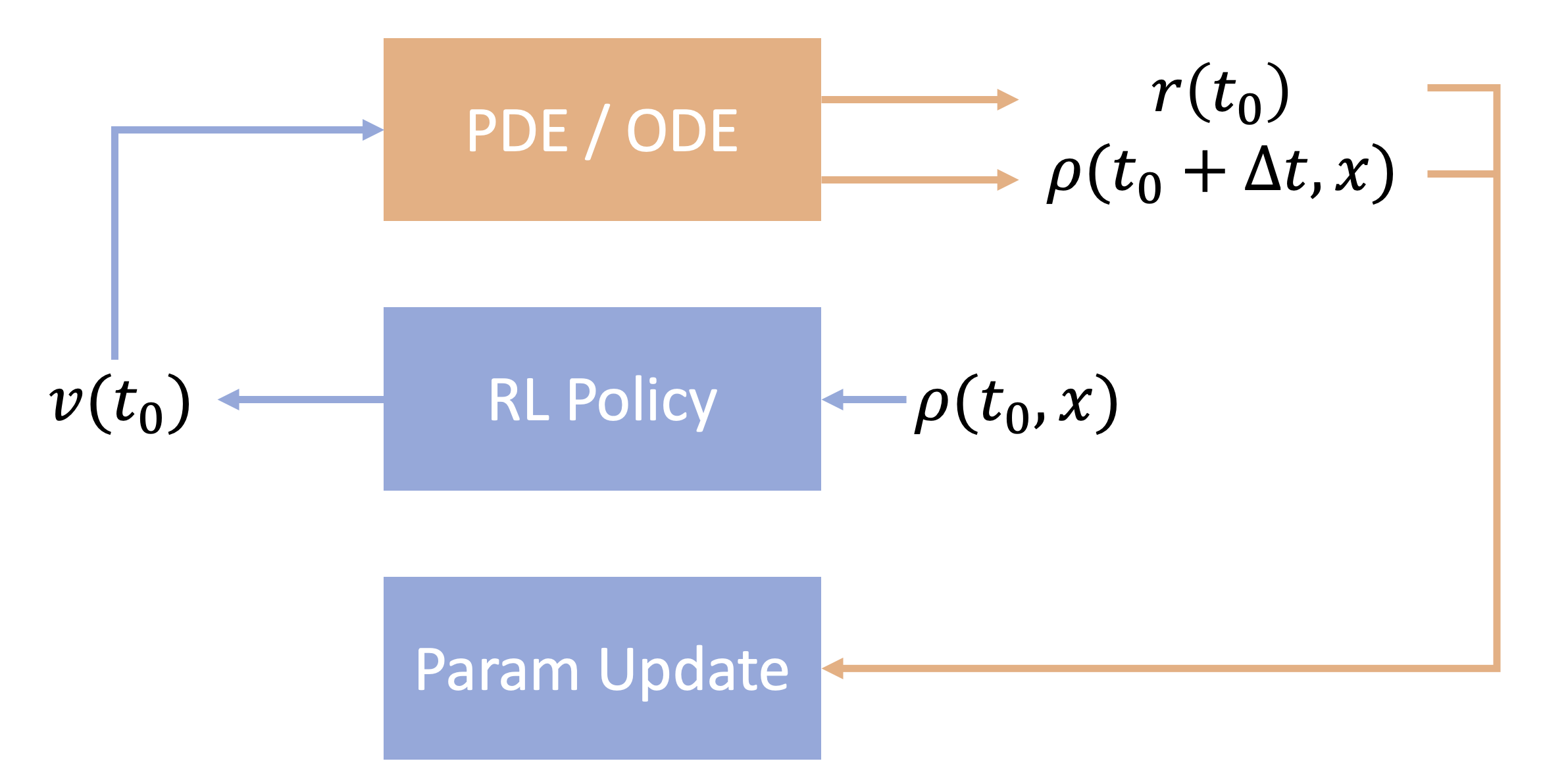}
    \caption{Control loop of the proposed RL-based Adaptive Controller.}
    \label{fig:control_loop}
\end{figure}
In this paper, we determine the control $V$ using the reinforcement learning method. We construct an adaptive controller based on a PDE-ODE model with the RL framework. The RL algorithm determines \textit{an optimal control policy} that adjusts the AVs speeds in real-time, with the objective of optimizing traffic flow with respect to predefined metrics (see \eqref{E:reward} and \eqref{E:optimal_policy}); Figure \ref{fig:control_loop} illustrates the general structure of this paper. 
\subsection{Markov Decision Process Formulation}
Theorem \ref{T:existence} proves the existence of the solution to the PDE-ODE problem \eqref{E:conservation}-\eqref{E:initial} for any generic control $V$. In fact,  \cite{garavello2020multiscale}  proved the existence of open-loop controls with bounded variations and later, in \cite{liard2020optimal} and \cite{liard2021pde} the existence of an optimal solution (even if may not be unique) was proven  
for a specific choice of cost functional corresponding to the fuel consumption. 
To frame our control problem within the RL paradigm, we define a Markov Decision Process (MDP). In the implementation of the RL scheme, we will discretize the $[0, T] \times [0, L]$ space both for the MDP as well as for the numerical scheme of the PDE-ODE problem presented in Section \ref{S:numerical}. As a rule of thumb, the discretization of the space should ensure that the total cell volume is not higher than the square of the maximum hidden state to ensure convergence and to elevate the curse of dimensionality. However, to explain the RL scheme in a more general setting, we will keep the notations in continuous state space for the rest of this section. \\
\textbf{State Space:} The state space for any fixed time $\hat t  \in [0, T]$ is defined as $\mathcal O(\hat t) \Def \set{\rho(\hat t, x) : x \in [0, L]}$. \\
\textbf{Action Space:} The actions for each fixed time $\hat t$ are the admissible set $\mathcal{A}(\hat t) \Def [0, V_{\max}]$ that can be assigned to the AV at any time step $\hat t$. In other words, at a fixed time $\hat t$, the piecewise constant control $V(\hat t) \in \mathcal A(\hat t)$. \\
     \textbf{Reward Function:} The reward function $\mathbf r$ is a composite measure that includes the Minimum Flux, the instant speed of the AVs, and the minimum deviation of global speed across the traffic flow. Mathematically, for any $(o(t), a(t)) \in \mathcal O(t) \times \mathcal A(t)$, the reward function $\mathbf r(t) = \mathbf r(o(t), a(t))$ is defined by 
  \begin{equation}\label{E:reward}
           \mathbf r(t) \Def w_1 \Phi(t) + w_2 V(t) - w_3 \totvar{v(t, \cdot)}{[0, L]}
  \end{equation}
  where for any $t \in [0, T] $ 
  \begin{equation*} \Phi (t) \Def \min \set{f(\rho(t, x)): x \in [0, L]}\end{equation*}  which is the minimum of the flux function, $V(t)$ is the AV speed at time $t$, and $\totvar{F(\cdot)}{\Omega}$ is the total variation of a function $F$ over a set $\Omega$. In fact, the last term implies the tendency to define the sequence of controls with minimum total variation as this will contribute to the stability of the numerical scheme for solving the PDE-ODE problem. 
  Here, $w_i$ for $i = 1,2, 3$ are corresponding weights.
  
In general, the objective of the RL agent is to update a policy $\pi(a(t) \mid o(t))$ such that

\begin{equation}\label{E:optimal_policy}
\pi^* = \underset{\pi}{\text{arg max}} \, \EE_{\tau \sim P_\pi} \left[\mathcal R(\tau) \right].
\end{equation}
 Here, $P_\pi(o(t), a(t))$ is the state and state-action marginals of the trajectory distribution that is induced by the policy $\pi(a(t) \mid o(t))$. The MDP trajectory and its corresponding return function are defined by 
 \begin{equation*}
 \tau \Def \set{(o(t),a(t)) \in \mathcal O(t) \times \mathcal A(t) |t\in [0, T]} ,
 \end{equation*} 
 and
\begin{equation*} 
\mathcal R(\tau) \Def \sum_{t\in [0,T]} \mathbf r(o(t), a(t)),\end{equation*}
respectively.

\subsection{Policy Parameter Update}
The RL policy parameters are updated through interactions with the PDE-ODE model. At each time $t$, the agent observes $\rho(t, x) \in \mathcal O(t)$, takes an action $V(t) \in \mathcal A(t)$ with respect to the current policy $\pi$ which follows by the reward $\mathcal R(\tau)$. 

The target optimal policy $\pi^*$ in this paper is adapted as $\pi_{\theta}$, a parameterized policy defined by a set of parameters $\theta$, and the update rule is derived from the policy gradient method. In particular, the update at each iteration $k+ 1$ can be expressed as:
\begin{equation}
\theta_{k+1} = \theta_k + \eta_{actor} \nabla_\theta J(\theta_k),
\end{equation}

where $\eta_{actor}$ is the learning rate and $J(\theta)\Def \EE_{\tau \sim \pi_\theta}[\mathcal R(\tau)]$, i.e., the expected return when following policy $\pi_\theta$. The gradient of the performance, $\nabla_\theta J(\theta)$, is estimated using the Actor-Critic approach, where the 'Actor' updates the policy in the direction of higher reward, and the 'Critic' estimates the value function, which is used to reduce the variance of the gradient estimate.

The Actor-Critic algorithm can be further detailed as follows:\\
\textbf{Actor Update:} The policy parameters are updated by taking steps in the direction of the gradient of the log-probability of the taken actions, weighted by the advantage function $A(t)$, which indicates how much better an action is compared to the average:
    \begin{equation}
    \theta_{k+1} = \theta_k + \eta_{actor} \mathbb{E}\left[\nabla_\theta \log \pi_\theta(V(t)|\rho(t, x)) A(t)\right].
    \end{equation} 
\textbf{Critic Update:} The value function parameters $\phi$ are updated on a predefined collection $\mathcal T = \set{s_\circ, \cdots, s_M}$ of times to minimize the difference between the predicted value and the actual return. More precisely, for any $\hat t \in \mathcal T$, the parameter updates are defined as:
    \begin{equation}
    \phi_{k+1} = \phi_k - \eta_{critic} \nabla_\phi (V_{\phi}(\mathcal O(\hat t)) - \mathcal R(\tau))^2, 
    \end{equation}
    where $\eta_{critic} $ is the learning rate for the Critic, and $V_{\phi}$ is the value function estimating the expected return given the state $\mathcal O(\cdot)$. 

By iteratively applying these updates, the RL agent refines its policy towards one that can effectively control the AVs to achieve the desired traffic flow characteristics.
\section{Numerical Results}\label{S:numerical}
To assess the efficacy of the adaptive speed controller for AVs in mixed autonomy traffic, we conducted a series of numerical experiments. These experiments were designed to emulate realistic traffic congestion scenarios, allowing us to evaluate the performance of the proposed method under various traffic conditions. The scenarios were constructed to reflect common congestion patterns observed in highway traffic, including peak-hour traffic flow, incidents-induced congestion, and congestion triggered by variable demand levels.

The performance of the adaptive speed controller was quantified using three primary metrics, which also served as the reward function for the reinforcement learning (RL) agent during training. These metrics are critical for understanding the impact of AVs on traffic dynamics and for evaluating the benefits of the proposed control strategy. Below we describe each metric in detail and present the results obtained from the numerical experiments.

\textbf{Minimum Flux:} The minimum flux metric measures the real-time lowest rate of traffic flow during the simulation period. This metric is indicative of the system ability to maintain traffic movement and avoid traffic jams, which are often the most critical periods during congestion. A higher minimum flux value suggests that the system is more effective at preventing severe congestion and ensuring smoother traffic flow.

In our simulations, we observed that the adaptive speed controller was able to improve the minimum flux by 15\% compared to the baseline scenario (without AV control). This improvement demonstrates the capability to mitigate the formation of traffic bottlenecks and maintain a higher rate of flow even under heavy traffic conditions.

\textbf{Instant Speed:} The current speed metric represents the average speed of the controlled AV (ego speed) and consequently the average speed of all vehicles in the upstream traffic (global speed). This metric provides insight into the energy efficiency of the traffic system and the comfort of the driving experience.

Our results indicate that the average ego speed decreased by 27\%, while the average global speed saw an enhancement of 17\% with the implementation of the adaptive speed controller, suggesting that the AV facilitated a faster average speed for the overall human-driven vehicles.

\textbf{Total Variation of Traffic Speed:} This metric measures the variability of speeds across the traffic. Lower speed deviation is generally associated with more stable and consistent traffic flow, which can lead to reduced fuel consumption and lower emissions.

The adaptive speed controller achieved a 35\% reduction in speed deviation, indicating a more uniform traffic flow. This reduction in speed variability can be attributed to the controller's ability to adapt to changing traffic conditions and to modulate the speed of AV to prevent shockwave formation in the traffic stream.

\begin{figure}
    \centering
    \includegraphics[width=0.6\linewidth]{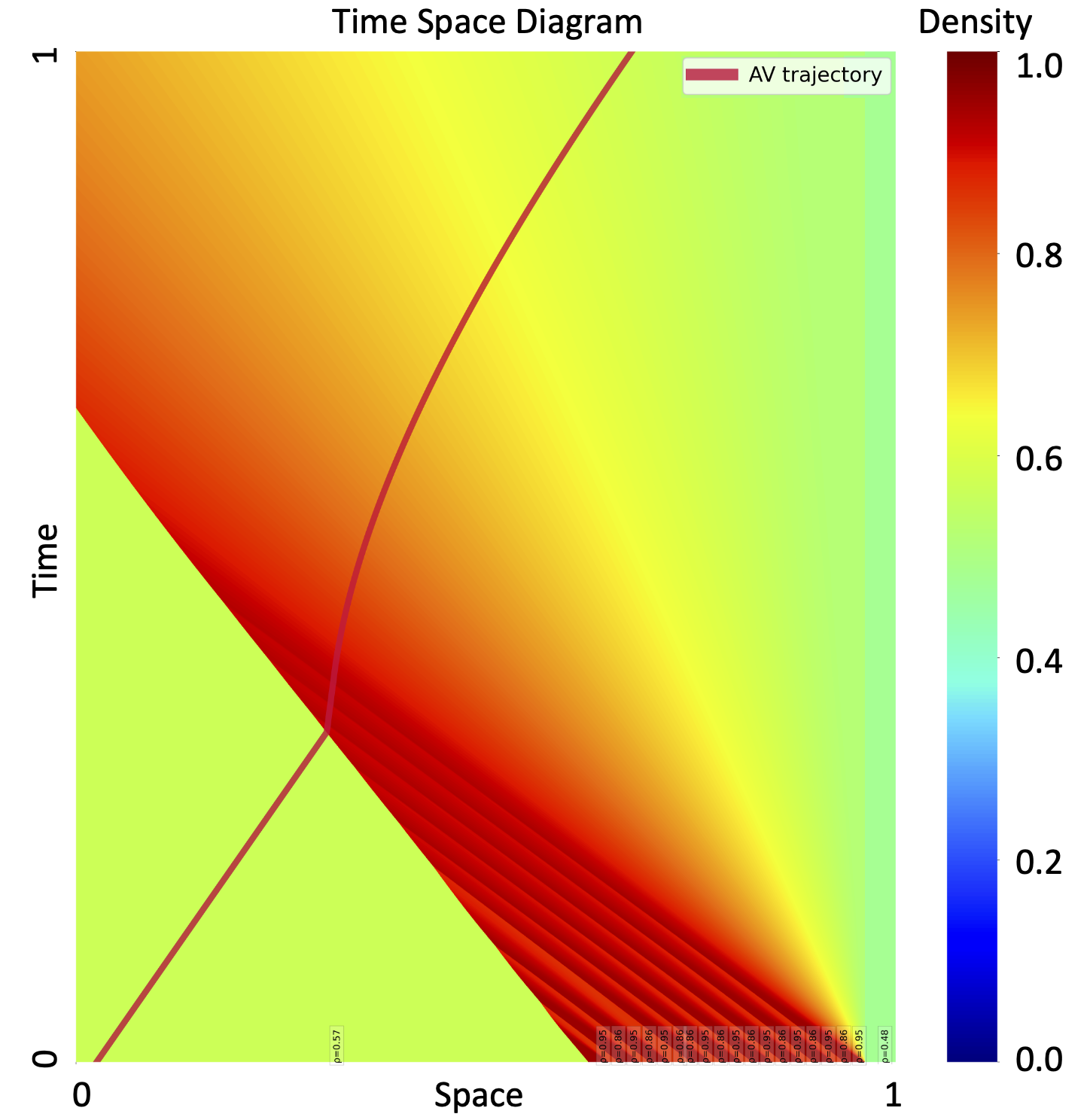}
    \caption{Benchmark scenario: Stop and go waves.}
    \label{fig:tsd_benchmark}
\end{figure}

\begin{figure}
    \centering
    \includegraphics[width=0.6\linewidth]{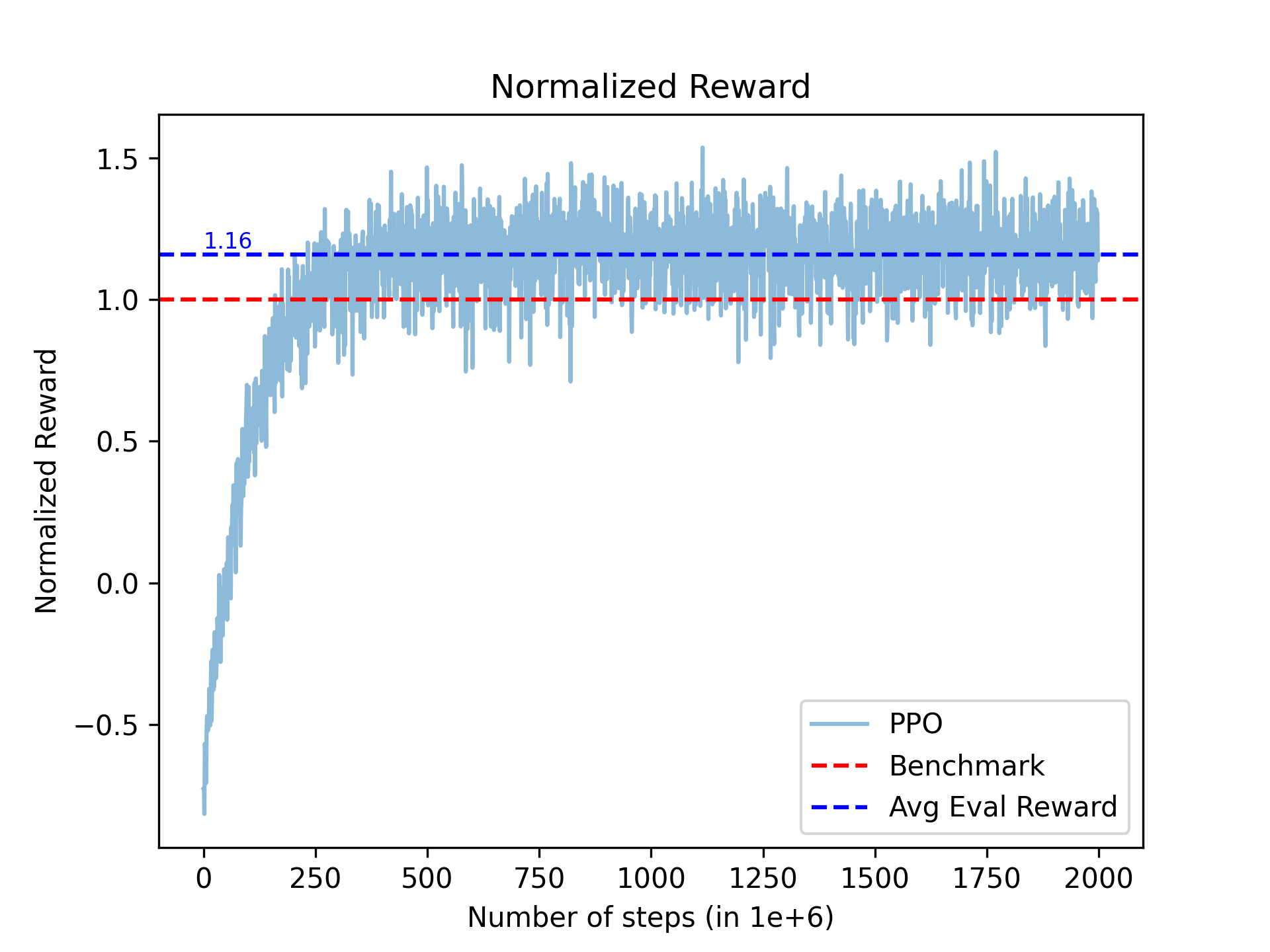}
    \caption{Learning curve of the PPO algorithm. 16\% of normalized reward improvement observed in evaluation.}
    \label{fig:lc}
\end{figure}

To quantify the impact of AVs acting as adaptive controllers in a mixed traffic environment, we employed the Proximal Policy Optimization (PPO)\cite{schulman2017proximal} algorithm, which introduces a novel objective function that employs a surrogate maximization problem with clipped probability ratios, effectively balancing the exploration of new strategies against the exploitation of known beneficial actions. This mechanism ensures that updates to the policy parameters are both substantial and stable, preventing detrimental large shifts that could arise from highly advantageous outlier actions. A shockwave scenario, as shown in Figure \ref{fig:tsd_benchmark}, was constructed as the testing benchmark. The learning curve depicted in Figure \ref{fig:lc} illustrates the progression of the PPO algorithm performance, highlighting a 16\% improvement in normalized reward, which underscores the algorithm capability to iteratively refine the AV control strategies for improved traffic management. A comparative analysis of the proposed system was conducted against the benchmark scenarios under various initial conditions. A comprehensive comparison of numerical results is shown in \textbf{Table \ref{tab:result}}.

\begin{figure}
    \centering
    \includegraphics[width=0.6\linewidth]{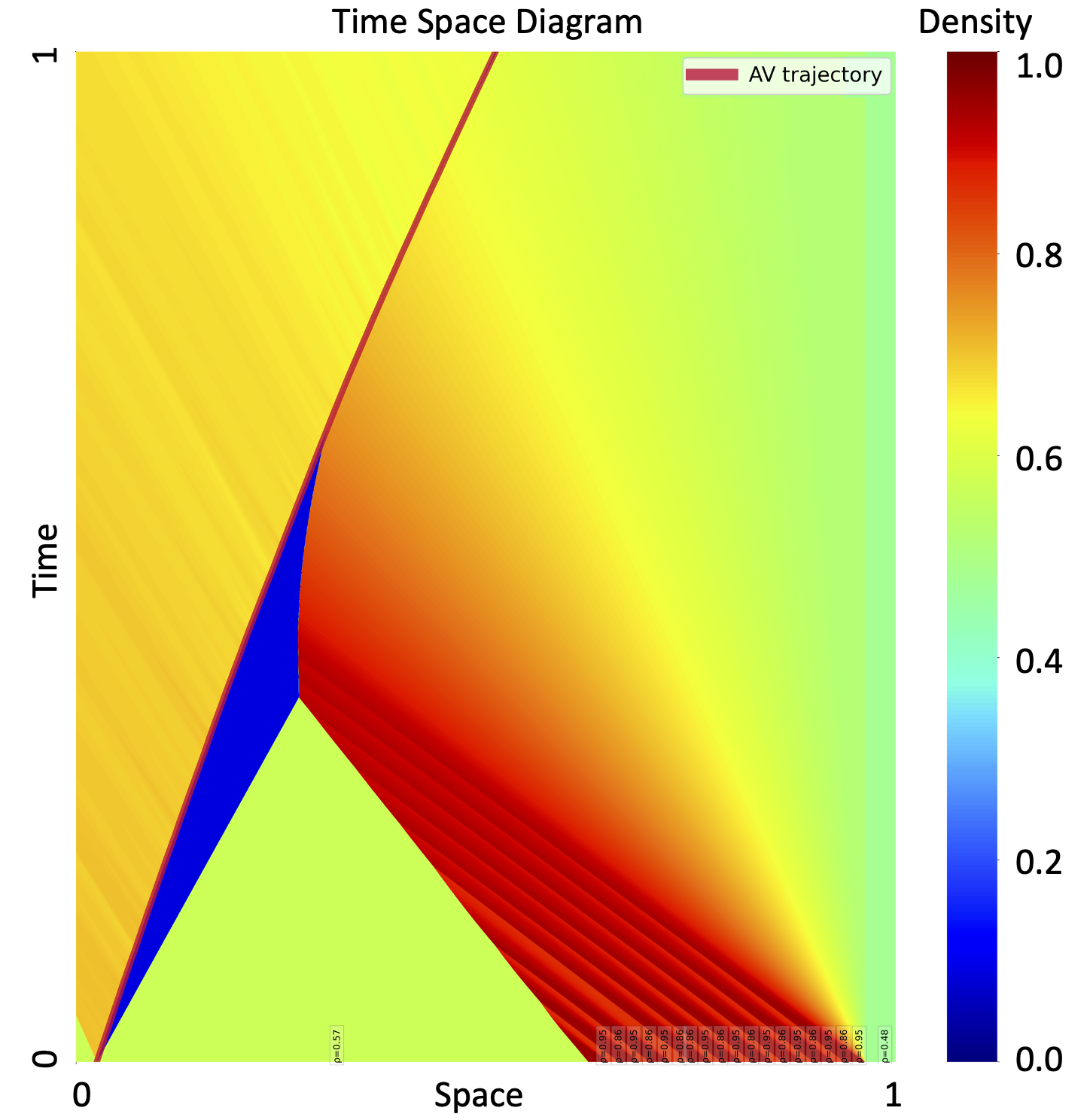}
    \caption{The reward weights of the model are $w_1,w_2,w_3=0.2,0.3,0.5$. The controlled vehicle tends to create the low-density area to neutralize the backward propagation of the shockwave. }
    \label{fig:tsd_0}
\end{figure}

\begin{figure}
    \centering
    \includegraphics[width=0.6\linewidth]{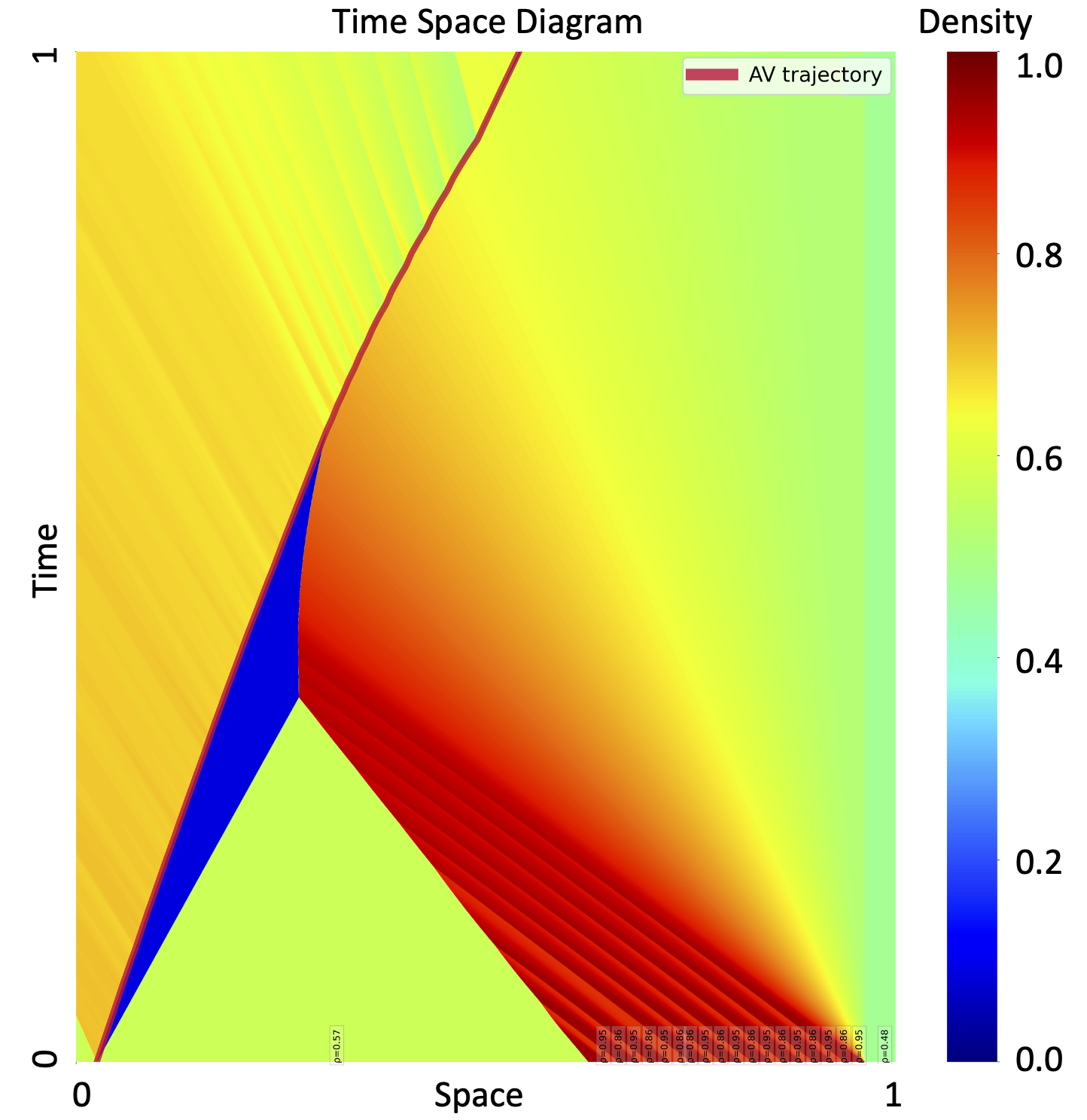}
    \caption{\small{The reward weights are adjusted to $w_1,w_2,w_3=0.2,0.5,0.3$ to emphasize the flux improvement. The major difference could be observed when the controlled AV driving through the rare fraction, where more oscillation of speed are observed behind the controlled AV.}}
    \label{fig:tsd_2}
\end{figure}

\begin{figure}
    \centering
    \includegraphics[width=0.6\linewidth]{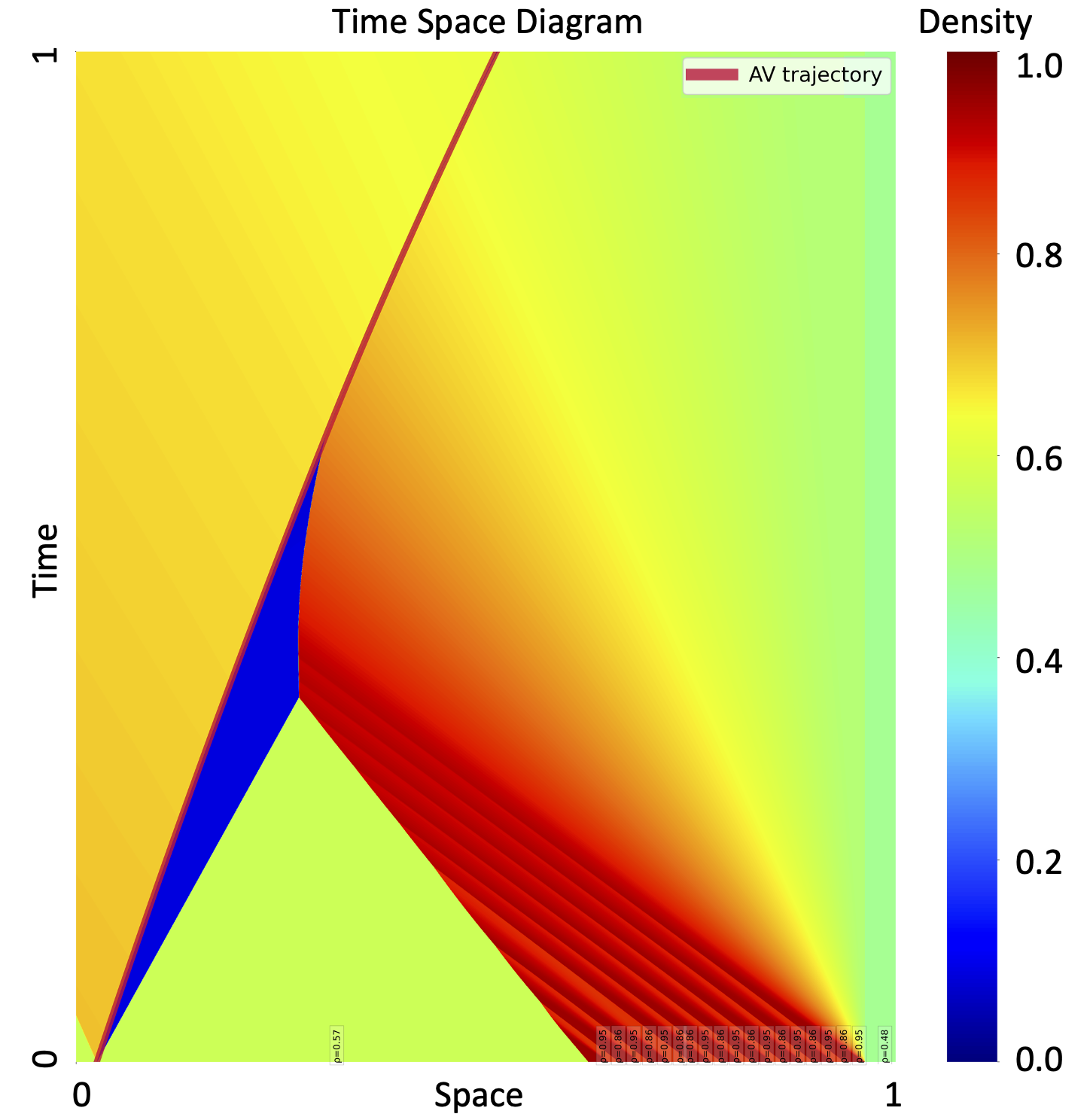}
    \caption{\small{The reward weights are adjusted to $w_1,w_2,w_3=0.1,0,0.9$. A more smoothed traffic flow is obtained by using the deviation-emphasized reward. The AV tends to avoid drastic speed changes to achieve a homogeneous traffic flow after the shockwave.}}
    \label{fig:tsd_3}
\end{figure}

The series of time-space diagrams from Figure \ref{fig:tsd_0} to Figure \ref{fig:tsd_3} elucidate the impact of varying reward weight configurations on the traffic flow, demonstrating the flexibility of the AV adaptive control system. In Figure \ref{fig:tsd_0}, we observe the AV ability to create a low-density area downstream, effectively dissipating the backward propagation of a shockwave, under a balanced reward structure. Altering the reward weights to favor flux improvement, as shown in Figure \ref{fig:tsd_2}, results in a distinctive pattern of speed oscillations, especially as the AV navigates through areas of rarefaction.

\begin{figure}
    \centering
    \includegraphics[width=1\linewidth]{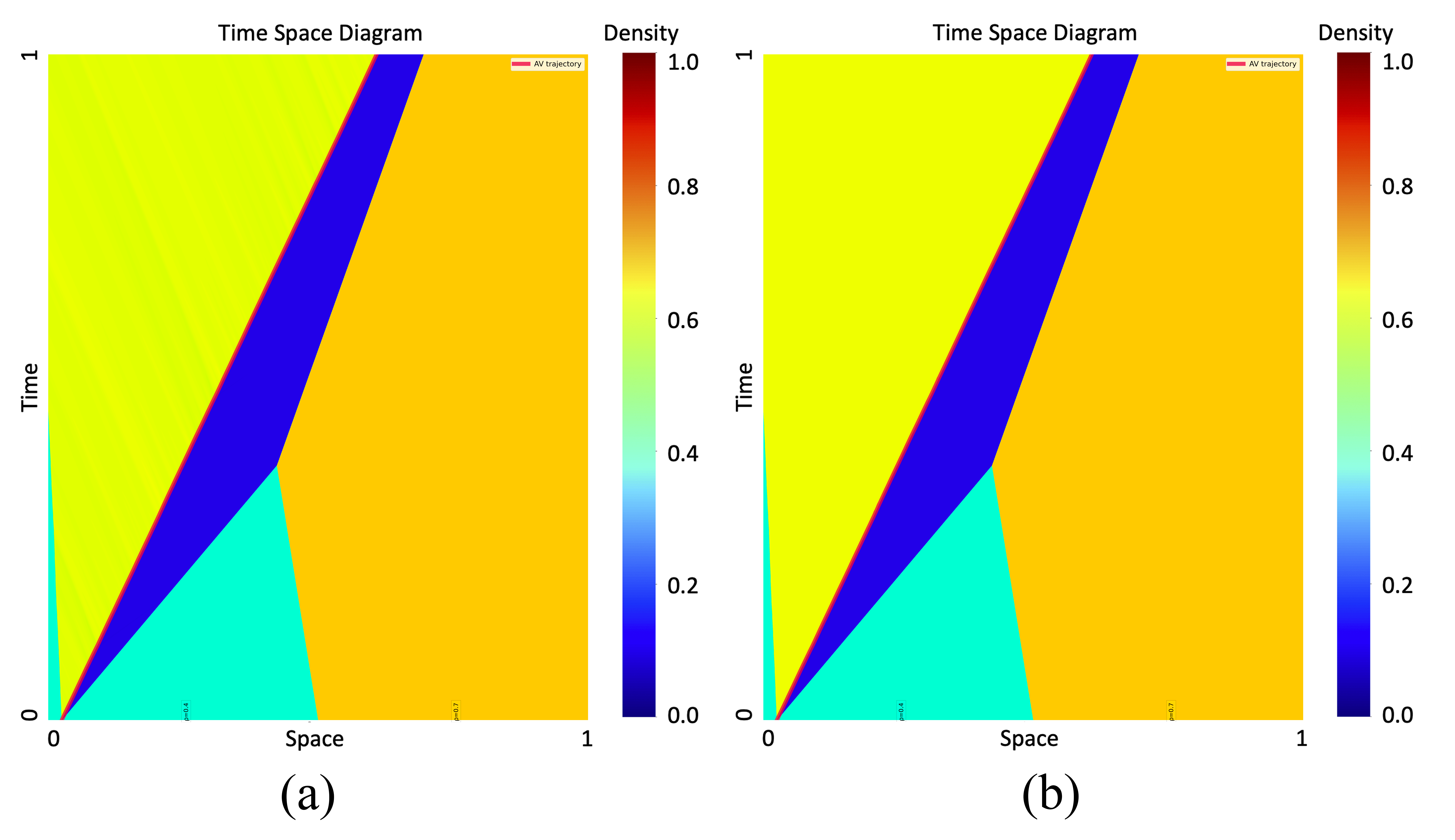}
    \caption{\small{The bottleneck scenario controlled by: (a) PPO-trained stochastic policy; (b) Deterministic policy using the mean of the distribution as the action. By comparing the two policies, we note that the oscillation in the speed profile can be attributed to the stochastic nature of RL output; i.e., removing the uncertainty eliminates the oscillations.}
    }
    \label{fig:tsd}
\end{figure}

Figure \ref{fig:tsd_3} presents a scenario where the reward system is adjusted to minimize speed deviations, leading to a more uniform traffic flow. The AV adopts a strategy that shuns abrupt speed changes, promoting a consistent traffic density and reducing the likelihood of shockwave formation. Conversely, a reward configuration that prioritizes the AV speed is investigated, which, while still capable of disrupting shockwave continuity through the creation of low-density zones, also introduces new disturbances in the traffic pattern due to the AV sudden speed variations.\\
These remarks serve to highlight the nuanced relationship between the chosen reward configurations and the resulting traffic flow dynamics. They provide a compelling demonstration of how reinforcement learning can be harnessed to develop control policies for AVs, enabling them to adapt to a range of traffic conditions and optimization goals. See also Figure \ref{fig:tsd} for some highlights on the connection between the oscillation in the optimal solution and the random nature of the RL policy.

\begin{table}[!htp]\centering
\caption{Numerical Simulation Results}\label{tab:result}
\scriptsize
\begin{tabular}{lrrrrrr}\toprule
$[w_1, w_2, w_3]$ &Avg. Flux & Ego Speed & Avg. Speed & Avg. Deviation \\\midrule
No Control &0.2055 &0.3318 &0.3318 &0.1220 \\
$[0.2, 0.5, 0.3]$ &0.2381 &0.2440 &0.3912 &0.0429 \\
$[0.2, 0.3, 0.5]$ &0.2433 &0.2679 &0.4183 &0.0582 \\
$[0.1, 0, 0.9]$ &0.2346 &0.2105 &0.3714 &0.0317 \\
$[0.9, 0, 0.1]$ &0.1855 &0.2982 &0.2461 &0.0952 \\

\bottomrule
\end{tabular}
\end{table}



\section{Conclusion}\label{sec:conclusion}
This paper has introduced an innovative approach to traffic control by leveraging the capabilities of autonomous vehicles (AVs) as dynamic actuators within mixed traffic flows. By employing a mathematical framework that combines PDE and ODE, we have modeled the dynamics of traffic flow and the influence of AVs therein. The core of our methodology is a Reinforcement Learning (RL) algorithm, specifically designed to adaptively adjust the speed of AVs to optimize traffic conditions.

Our results demonstrate that the proposed RL-based adaptive speed controller can effectively mitigate traffic congestion by creating and manipulating low-density areas in the traffic flow, which in turn neutralizes shockwave propagation. Through numerical simulations, we have shown that the controller can significantly improve traffic metrics such as flux, average speed, and speed deviation. The versatility of the controller was further highlighted by its performance under various reward structures, which allowed for a tailored approach to either collective traffic flow optimization or individual AV performance.

The implications of this research are twofold. Firstly, it provides a practical framework for the deployment of AVs as a means to enhance traffic efficiency, which could be a pivotal step towards the realization of smart transportation systems. Secondly, it contributes to the theoretical understanding of traffic dynamics and control in the era of autonomous driving technology.

Future work will focus on expanding the scalability of the control method to larger traffic networks, exploring the impact of different levels of AV penetration in traffic, and integrating real-world traffic data to further refine the model. We will also aim to integrate explicit safety protocols into the adaptive speed controller. This will include measures such as safe distance keeping, emergency maneuver support, and the prediction of human driver behaviors to ensure a traffic environment that is not only efficient but also maximally safe for all participants. Additionally, we aim to investigate the socio-economic effects of implementing such AV-based traffic control systems, considering factors such as user acceptance, policy implications, and environmental impact.









\bibliographystyle{plain}
\bibliography{references}
\end{document}